\documentclass[journal, a4paper]{IEEEtran}

\usepackage{graphicx}
\usepackage{amssymb}
\usepackage{amsmath}
\usepackage[caption=false,font=footnotesize]{subfig}
\usepackage{multirow}
\usepackage{booktabs}
\usepackage{hyperref}

\usepackage{lineno}
\usepackage{pbalance}

\newcommand{\refFig}[1]{Figure~\ref{fig:#1}}

\begin{document}

\title{An advanced scheme for queue management in TCP/IP networks}

\author{\IEEEauthorblockN{Abderrahmane Boudi \IEEEauthorrefmark{1}, Malik Loudini \IEEEauthorrefmark{1}}\\
\IEEEauthorblockA{\IEEEauthorrefmark{1} \'{E}cole nationale Sup\'{e}rieure d'Informatique (ESI), Algiers, Algeria, \{a\_boudi, m\_loudini\}@esi.dz}
}

\maketitle

\begin{abstract}
Active Queue Management (AQM) is a key congestion control scheme that aims to find a balance between keeping high link utilization, minimizing queuing delays, and ensuring a fair share of the bandwidth between the competing flows.
Traditional AQM mechanisms use only information that is present at the intermediate nodes (routers).
They do not take into account the particularities of the flows composing the traffic.
In this paper, we make use of a mechanism, called Explicit RTT Notification (ERN), that shares with routers information about the Round Trip Times (RTTs) of the flows.
We propose a new fuzzy logic based AQM controller that relies on the RTTs of the flows to improve fairness between them.
The performances of the new proposed method, FuzzyRTT, is examined and compared to existing schemes via simulation experiments.
\end{abstract}


\section{Introduction}
\label{sec:introduction}

Active queue management (AQM) mechanisms help maximize utilization while minimizing the queuing delays of routers in TCP-IP networks.
They do so by sending congestion notifications to the flows when the congestion is not yet severe.
This type of congestion control, unlike TCP's congestion control, is done from within the network.
One of the first AQM schemes is RED~\cite{Floyd1993}.
It was proposed by Floyd and Jacobson to avoid the flow synchronization problem.

Since RED, a great number of AQM mechanisms have been proposed.
Each one of them proposes to enhance queue control in some way or another.
Works such as~\cite{Feng1999,Athuraliya2001,RongPan2000} have changed the way congestion measurement and detection are done.
While others, such as~\cite{Hollot2001,Fengyuan2002,Yanfie2003,Chrysostomou2009}, borrowed concepts from feedback control theory to introduce robustness and stability in the study of AQM schemes.
Works such as~\cite{Lin1997,RongPan2000,Lu2014} have concentrated their effort in improving the fairness of AQM schemes.
There are also works such as~\cite{Nichols2012,Xiong2010,Kahe2013} that tried to relieve the networks operators from tuning the AQM schemes.
Finally, works such as~\cite{Katabi2002,Xia2008,Wang2013} have proposed hybrid AQM schemes in the sense that their mechanism of congestion controls is composed of an end-to-end protocol running at the edge of the network coupled with an AQM scheme that operates from within the network.

In this paper, FuzzyRTT is presented; a new method for controlling the queue during times of congestion.
We propose to take advantage of the RTT information to improve the distribution of the bandwidth between flows, and thus increasing the fairness among them.
The novelty of the proposed solution lies in the fact that it only uses the RTT information to greatly improve the fairness among flows.
Plus, unlike other solutions, FuzzyRTT is a pure AQM solution.
Schemes such as~\cite{Katabi2002,Xia2008,Wang2013} have proposed congestion control protocols operating at the end nodes coupled with AQM schemes operating at the router level.
While FuzzyRTT does not require any specific transport protocol operating at end nodes.
To the best knowledge of the authors, only a few studies~\cite{Hoshihara2006,Shyu2004,Grazia2015,Casoni2017} took the RTTs of the flows into account when implementing their AQM schemes.
Up to now, existing fuzzy logic based AQM schemes have not taken into account the RTT of the flows when building their controllers.
With FuzzyRTT, we are proposing a fuzzy logic controller (FLC) that involves the RTTs of the flows in its decision process.
Another core aspect of FuzzyRTT is its statelessness, it does not need to maintain the state of the flows traversing the link.
This is why throughout this paper, as was done in~\cite{Casoni2017}, we shall suppose that the RTT information is appended to each packet of long lived FTP flows.
Nonetheless, we shall explain how this can be done in Section~\ref{sec:deployment}.

The remainder of the paper is organized as follows.
Section~\ref{sec:background} contains a brief overview of the related works.
In Section~\ref{sec:design}, the design guidelines of the new AQM scheme FuzzyRTT are drawn\@.
In Section~\ref{sec:setup}, we define the simulation setup.
Section~\ref{sec:results} discusses the performance of the proposed scheme through a set of simulation exercises, also comparing with other well-known AQM techniques.
In Section~\ref{sec:deployment}, we address the deployment issues.
Finally, we draw our conclusions and future works in Section~\ref{sec:conclusion}.

\section{Background and related work}
\label{sec:background}

Even after nearly three decades of active research, there are still fresh proposition of AQMs mechanisms that tries to tackle the congestion problem in TCP-IP networks.
In this section, we will introduce the latest and most important works that are relevant to  our case study.
Given the huge number of proposed solutions, several surveys studying AQMs were conducted during these years.
Maybe the most recent are Adam's survey~\cite{Adams2013} on Active Queue Management and Abbas's~\cite{Abbas2016} on the fairness of AQM schemes.

The oldest, most widespread, and studied AQM scheme is RED~\cite{Floyd1993}.
It was proposed by Floyd and Jacobson to counter the flow synchronization problem which is mainly due to the way passive queue management mechanisms control the queues.
Indeed, passive mechanisms begin to drop packets only when the queue becomes full be it from the tail, front, or randomly from within the queue.
When the queue of the bottleneck link becomes full, it drops a great number of packets from different flows at the same time, which leads to the synchronization of these flows.
When the flows are synchronized, the network behaves in a cyclic manner.
Indeed, when the queue becomes full, the router drops packets from most of the flows composing the traffic.
All the notified flows will reduce their sending rates at the same time which leads to an under utilization of the link.
From there, the flows will, more or less, uniformly increase their congestion windows, which will rapidly lead to full utilization of the link, then large queuing times, then finally to congestion again.

The idea behind RED, or any other AQM scheme, is to drop packets at the early stages of the congestion.
Doing so prevents the synchronization problem and therefore increases the utilization of the link while reducing queuing times.
Indeed, notifying flows at the early stages of the congestion will spread the notifications in time which means that different flows will be notified at different times.
Therefore, the queue will not build-up and the reduction of the rates of the flows will not happen at the same time.
This is why the Internet Engineering Task Force (IETF) recommends the use of AQM in internet routers~\cite{Baker2015} and RED is included in most of routers.
Unfortunately, it was shown to be hard to tune the parameters of the RED scheme~\cite{Feng1999-2}.

Every AQM scheme is composed of two different parts.
The first part is concerned about detecting the congestion while the second part is about notifying the flows about the occurrence of congestion.
Works such as~\cite{Floyd1993,Lin1997,Feng1999-2,Jamali2013} detect the congestion by monitoring the average length of the queue or its variation in time.
Misra et al.~\cite{Misra2000} have proposed a mathematical model based on TCP to catch the dynamics of AQM\@.
Many works, such as~\cite{Hollot2001,Yanfie2003,Xiong2010}, have used Misra's model to propose AQM schemes based on control theory.
These works rely mainly on the queue length and its variation to detect the congestion.
Other works, such as~\cite{Feng1999,Athuraliya2001,Gao2003}, argued that using the queue length is not the right metric to detect congestion.
They used instead the difference between the incoming and outgoing rates.
In~\cite{Nichols2012}, the authors use the sojourn time of the packets to detect congestion.
They argued that an AQM using the sojourn time instead of the queue length would prove to be more robust against changes in the capacity of the link.

There are mainly two ways of notifying the flows about the emergence of congestion, implicit or explicit congestion notifications.
Notifying implicitly means that the router needs to drop packets in order to notify the flows about the congestion.
The notified flow reduces its congestion window when it receives the same acknowledgment three times in a row or when a timeout occurs.
Explicit notification consists in informing the flows about the congestion without dropping packets.
Explicit Congestion Notification (ECN)~\cite{Ramakrishnan2001} was introduced in the TCP/IP header to robustly notify the end nodes to reduce their sending rates.
Most of the works cited above support both techniques of notification.
This is because TCP handles both of missing packets and marked packets as being due to congestion in the network.
Luckie~\cite{Luckie2015} has recently proposed RECN (Really-ECN) that sends ICMP packets to hosts to advise them to reduce their rate.

Other works have argued that in order to maximize the benefits of AQM schemes, there should be an interplay between the end nodes and the routers.
In~\cite{Katabi2002}, Katabi et al.\ proposed eXplicit Control Protocol (XCP).
XCP is a transport protocol that runs in end nodes and in routers.
End nodes share with routers information about their sending rates;
and the routers advertise to the end nodes the preferred sending rate that should be used.
XCP ensures high utilization with a fast convergence towards a high level of fairness among flows.
Many works such as~\cite{Low2005,Zhang2005,Zhou2013} have studied and proposed enhancements to XCP;
but it was not largely deployed due to the significant changes it introduces in the functioning of end nodes and routers.
In~\cite{Xia2008}, Xia et al.\ proposed Variable-structure congestion Control Protocol (VCP).
The main idea behind VCP is the same as XCP, but unlike XCP, the authors use only the already existing bits of the TCP header.
Therefore, the standardization process of VCP would have been easier compared to XCP's\@.
Although, VCP, like XCP, requires a tedious deployment process.
UNO~\cite{Vasic2009} was proposed as trade-off between XCP and VCP\@.
Like XCP, UNO shares the RTT information with the routers, but it does so by only using the already existing ECN flags.
All these schemes suffer from a tedious deployment process, because they require to introduce significant changes in the protocols operating at the end nodes.

The fairness of an AQM scheme is determined by the way the bandwidth is shared among the competing flows.
In order to enforce fairness among flows, some AQM schemes, such as~\cite{Lin1997,Ramaswamy2007,Li2002,Xue2016}, maintain a list containing the characteristics of the active flows.
This kind of AQM schemes is called stateful.
Stateless AQM schemes are the ones that try to enforce fairness without the need to track the flows traversing the link.
Such schemes do not require any information about the competing flows to ensure some degree of fairness on the link~\cite{RongPan2000,Yeom2006,Lu2014}.
The last category is when the AQM requires only partial state.
Schemes such as~\cite{Parris1998,Wu-chunFeng2002,Mahajan2001,Ott1999} maintain partial state. 
While there are many ways on how to calculate the fairness of AQM schemes, most works on this field, rely on Jain's fairness index~\cite{Jain1984}.

A lot of effort has been put into estimating the properties of the flows composing the traffic.
Some works, such as~\cite{Aikat2003,Shakkottai2004,Phillipa2006,Landa2013,Fontugne2015,Ding2015}, have studied the distribution of RTTs of TCP connections.
From these works, it is clear that the flows in the Internet show a high heterogeneity of RTTs.
While other works, such as~\cite{But2005,Lance2005,Pei2009,Carra2010}, provided means to estimate the RTTs of the flows.
As was discussed in~\cite{Hollot2001,Floyd1991,Floyd1999}, the RTT plays an important role in the distribution of the bandwidth.
This is the reason why works such as~\cite{Hoshihara2006} proposes AQM schemes that rely on RTT measurements.

Fuzzy logic is a branch of Computational Intelligence (CI) that provides a mathematical tool for both collecting human knowledge and experience to deal with uncertainties in the control process.
It offers a mean to design controllers without the need to offer a mathematical model of the plant that is to be controlled.
It is used in a variety of different applications, and indeed AQM is one of them.
The first AQM scheme based on fuzzy logic is FLC-AQM~\cite{Fengyuan2002}.
FLC-AQM monitors the queue length and its variation in time, and provides the control signal which is the drop probability.
The performance of FLC-AQM was compared against the PI Controller~\cite{Hollot2001}.
Even after more than a decade, fuzzy logic based AQM schemes are still being studied~\cite{Liu2013,Revathi2011,Chrysostomou2009,Hosseini2009,Zhou2009,Liu2008,HadjadjAoul2007}.

\section{FuzzyRTT Controller design}
\label{sec:design}

Most of the time, AQM schemes maintain a drop probability that is applicable to all packets passing through the router.
Thus, when a packet reaches the router, the AQM scheme uses the drop probability as a mean to decide if it drops the packet or not.
Therefore, there are two different locations of decision making.
The first one is the drop probability calculation and the second is whether or not to drop the packet.
The first part is the queue management part and the second is the scheduling part~\cite{Braden1998}.
The queue management part is mostly concerned by driving the queue to the desired target.
While the scheduling part is more concerned about individual flows, where its main concern is to notify the right flow.
These two mechanisms should be separated because while queue management goal is to maximize utilization and minimize delays, the scheduling goal is to increase fairness among competing flows.
The reason why most of the AQM schemes do not explicitly separate between the two mechanisms is due to the absence of information about the flows at the routers.
Given that, in this work, the information about flows' RTTs is available, in what follows we shall address these two mechanisms separately.

\subsection{Queue management}
Queue management mechanisms are aimed to control the congestion, therefore their main purpose is to detect the severity of the congestion and decide about the appropriate response to counter it.
In order to detect it, the more widespread technique is to monitor the queue length.
As mentioned above, some works use the actual queue length, where other use the weighted exponential moving average.
In this work, inspired by the CoDel technique, we use the minimum queue length experienced in one RTT\@.
Consequently, the proposed controller has two inputs signals, which are the error \(e(kT)\) and its rate of change \(\Delta e(kT)\).
\(e(kT)\) is characterized by
\begin{equation} \label{eq:error}
e(kT) = qLen_{\min}(kT) - qLen_{target}
\end{equation}
where \(qLen_{target}\) is the expected queue length, \(qLen_{\min}\) is the minimum queue length that the queue of the router experienced during the last interval~\(T\), and \(kT\) is the \(k^{th}\) sampling instant.
\(\Delta e(kT)\) is the rate of change of the error; it is characterized by
\begin{equation} \label{eq:DeltaError}
\Delta e(kT) = e(kT) - e((k-1)T)
\end{equation}

The only way to control the queue is to drop or mark the packets when the congestion occurs, thus the output signal of a controller should be the decision to drop or not a packet.
Calculating such a decision for each passing packet is an overkill~\cite{Hollot2001,Fengyuan2002}.
Therefore, the output of the controller usually consists in a drop probability applicable to all packets passing the router.
The drop probability is then updated with the interval~\(T\).
Given the fact that \(qLen_{\min}\) is the congestion signal which is tightly related to the RTT, the interval~\(T\) should also be somewhat related to the RTT of the flows.

FuzzyRTT is a two inputs single output FLC.
Multiple Inputs Single Output (MISO) controllers usually capture more accurately the dynamic state of the controlled system.

The basic structure of a fuzzy system includes three main components: 
(1) the fuzzification and defuzzification units contain the membership functions, which allow translating input crisp data into fuzzy values and fuzzy output to crisp data, respectively;
(2) the decision table contains a set of rules characterizing the control policy and goals;
(3) the scaling factors consist on numbers that help normalize the crisp inputs to fall in the interval of the membership functions, and also help the output crisp data to be scaled to the right magnitude.
In what follows, we shall address how FuzzyRTT implements each of these three components.

\subsubsection{Fuzzification and defuzzification}
\begin{figure*}[!t]
  \centering
  \subfloat[input variable \(e(kT)\)]{
    \includegraphics[width=3.5in]{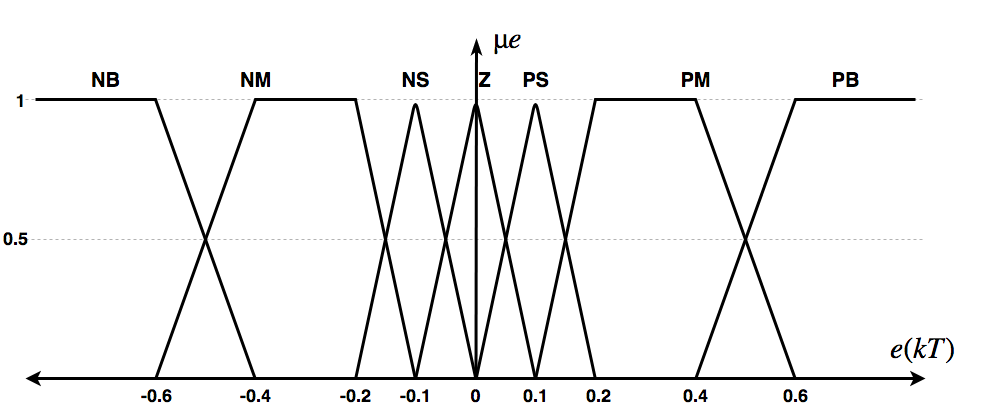}
\label{fig:memFunIn11}
  }
  \hfil
  \subfloat[input variable \(\Delta e(kT)\)]{
    \includegraphics[width=3.5in]{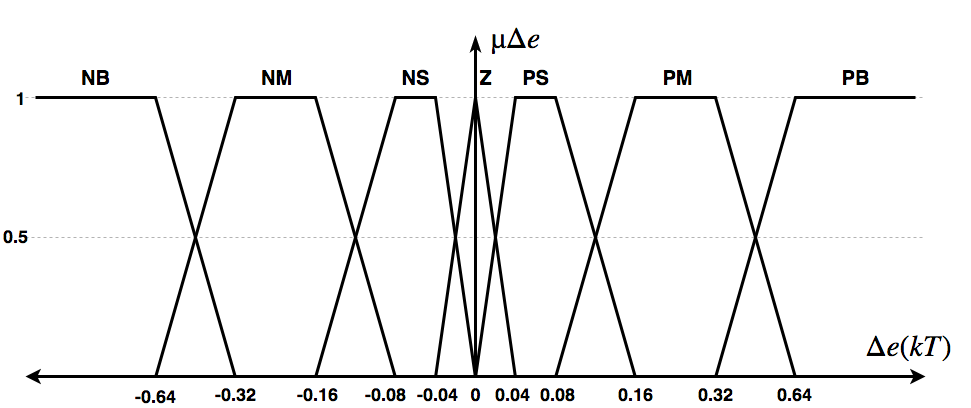}
\label{fig:memFunIn12}
  }
  \caption{Membership functions of the FLC input variables}
\label{fig:memFunIn1}
\end{figure*}

\begin{figure}[tb]
  \centering
  \includegraphics[width=3.5in]{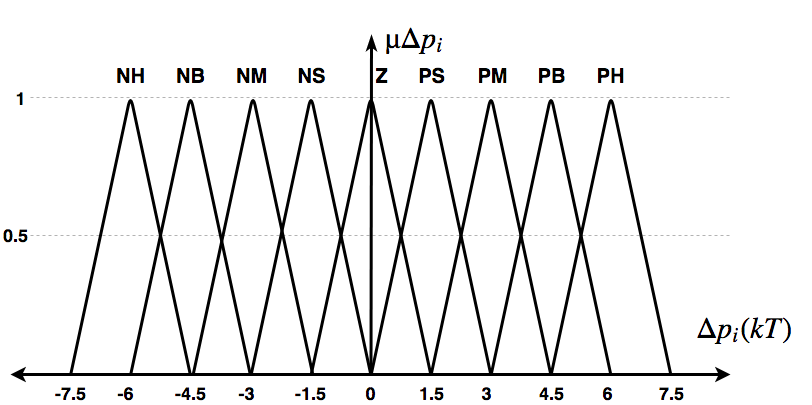}
  \caption{Membership functions of the FLC output variable \(\Delta p_{i}(kT)\)}
\label{fig:memFunOut1}
\end{figure}

One of the main aspects that characterizes the membership functions is the number of fuzzy terms and their shapes.
A large number of fuzzy terms help to more accurately catch the dynamics of the system;
but it increases the computation time and the complexity to the rule base.
FuzzyRTT has two inputs, the error \(e(kT)\) and its rate of change \(\Delta e(kT)\).
Both of the membership functions of these two inputs have seven fuzzy term sets, as shown in Fig.~\ref{fig:memFunIn1}.
The triangular and trapezoid shapes were chosen to reduce the computation complexity.
The number of fuzzy terms and the their shapes are initially chosen by the controller's designers.
Then, the fuzzy terms are refined according to the results of the carried experiments.
These choices represent a trade-off between catching the dynamics of the system and reducing the complexity of the controller.
The seven fuzzy term sets are negative big (NB), negative medium (NM), negative small (NS), zero (Z), positive small (PS), positive medium (PM), and positive big (PB).
It should be noted that the membership function of both inputs are normalized.
As shown on Fig.~\ref{fig:memFunOut1}, the output has nine fuzzy terms, it has negative huge (NH) and positive Huge (PH) in addition compared to the inputs.
The center of gravity is used to translate the fuzzy output to the actual value of the drop probability.
This defuzzification method is the most common method for defuzzification.
Finally, It also should be noted that FuzzyRTT is a Fuzzy-PI controller, in that the output signal is the necessary value that should be added to the drop probability to drive the system to the desired target.

\subsubsection{The rule base}
The rule base or the decision table reflects the particular view and experience of the designer.
We have used the well-known MacVicar-Whelan decision table. 
The decision table is presented in a tabular form (Table~\ref{table:decision}).

\begin{table}
\centering
\caption{FuzzyRTT rule table}
\label{table:decision}
\bgroup\
\def\arraystretch{1.5}%
\setlength{\tabcolsep}{.16667em}
\begin{tabular}{cc|ccccccc}
\toprule
\multicolumn{2}{c}{\multirow{2}{*}{\(\Delta p_{i}(kT)\)}} & \multicolumn{7}{c}{\(\Delta e(kT)\)}  \\ \cline{3-9}
\multicolumn{2}{c}{}                         & NB  & NM  & NS  & Z   & PS  & PM  & PB  \\ \midrule
\multirow{7}{*}{\(e(kT)\)}
                                & NB         & NH  & NH  & NB  & NB  & NM  & NS  & Z   \\ \cline{2-9}
                                & NM         & NH  & NH  & NB  & NM  & NS  & Z   & PS  \\ \cline{2-9}
                                & NS         & NB  & NB  & NM  & NS  & Z   & PS  & PM  \\ \cline{2-9}
                                & Z          & NB  & NM  & NS  & Z   & PS  & PM  & PB  \\ \cline{2-9}
                                & PS         & NM  & NS  & Z   & PS  & PM  & PB  & PB  \\ \cline{2-9}
                                & PM         & NS  & Z   & PS  & PM  & PB  & PH  & PH  \\ \cline{2-9}
                                & PB         & Z   & PS  & PM  & PB  & PB  & PH  & PH  \\ \bottomrule
\end{tabular}
\egroup\
\end{table}

\subsubsection{The scaling factors}
The scaling factors help size the input signal to the right magnitude.
They are an important factor of any FLC.
An improper choice during the selection of the scaling factors may lead to instability and oscillation of the controlled system.

In order to normalize the queue length error \(e(kT)\), the scaling factor \(SF_{i1}\) is set to
\begin{equation} \label{eq:sfi1}
  SF_{i1} = \dfrac{1}{qLen_{target}}
\end{equation}
where \(qLen_{target}\) is the target queue length.
It is clear, using Eq.\ref{eq:sfi1}, that the input signal will fall in the [-1 1] interval when the queue length is between 0 and \(2 * qLen_{target}\).
Thus, when the queue length is above \(2 * qLen_{target}\), the input signal is bounded by 1, which means that the error would be PB.

In order to have consistent results under several network configurations, \(\Delta e(kT)\) should adapt to the capacity of the link.
Thus, \(SF_{i2}\) is calculated so the maximum shift would be worth \(10 ms\) of queuing time.
Consequently, the scaling factor of the \(\Delta e(kT)\), it is calculated by
\begin{equation} \label{eq:sfi2_1}
  SF_{i2} = \dfrac{1}{C/(8*MSS)*0.01}
\end{equation}
where \(C\) is the capacity of the link in \(bps\), \(MSS\) is the maximum segment size, and \(0.01\) represent \(10 ms\) worth of queuing time.
When simplifying Eq.\ref{eq:sfi2_1}, it becomes
\begin{equation} \label{eq:sfi2_2}
  SF_{i2} = \dfrac{800*MSS}{C}
\end{equation}
Finally, the scaling factor of the output is calculated by
\begin{equation} \label{eq:sfo}
  SF_{o} = \log(N)
\end{equation}
where \(N\) is the number of active flows.
As discussed in~\cite{Feng1997}, the drop probability should take into account the number of active flows.
Therefore, in Eq.\ref{eq:sfo}, the logarithmic function was used to smooth the scaling factor value.
Also, by not using the logarithmic, we would have a large output, which will result in large oscillation and instability of the system.

\subsection{Scheduling}
When an outgoing interface of a router receives a packet, it forwards, marks, or drops the packet.
If that decision is made without considering the flow, but only aggregated information such as the queue length, this would mean, then, that the router is treating all packets equally.
Which will result in a negative impact on the fairness because different flows have different characteristics.

In this work, we differentiate between flows mainly by their RTTs.
Therefore, FuzzyRTT should take into account the RTT of each flow when deciding whether to drop or not a packet.
Having one controller and scaling the drop probability in accordance with the RTT of the flow is not a viable option to control the queue.
This is due to the fact that the FuzzyRTT controller is greatly bound to the interval of update~\(T\).
Thus, having only one controller is meaningless considering that the detection of the congestion is bound to the RTT of the flow.
While having one controller for each flow is also not a viable option due to the overhead of computation exerted on the router.
Thus, the best course of action would be to segment the flows into a fixed number of categories.
The RTTs of the flows vary from tens of milliseconds up to hundreds.
According to \cite{Kuhn2016}, the maximum value for the RTT can be set to \(560ms\).
Therefore, we chose to maintain five drop probabilities, thus we have five different categories.
As we think that this is a good trade off between minimizing computation overhead while offering a good segmentation of the universe of discourse.
The first category is for the flows with an RTT less than \(40ms\);
this RTT is doubled each time for the remaining categories.
Therefore, the second, third, fourth, and fifth categories have RTTs of \(80ms\), \(160ms\), \(320ms\), and \(640ms\), respectively.
The flows with RTTs greater that \(640ms\) belong to the fifth category.
These five drop probabilities are calculated using the FLC described above.
The only thing that changes is the update frequency of the drop probability and \(qLen_{\min}\).
Doing this, would help to adapt the control for each type flow instead of treating all the flows in the same way.

Each category has a distinct FuzzyRTT controller with an interval of update following the RTTs presented above.
Given the fact that the interval of update between each two adjacent categories is doubled each time, even if the two controller have the same input, it would take the double of the time for the controller with the higher delay to reach the same output as the one with the lower delay.
Therefore, the output scaling factor of categories 2, 3, 4, and 5 is multiplied by 2, 4, 8, and 16, respectively.
One concern when using different frequencies would be the starvation of flows with small delays.
Indeed, when the congestion is caused by flows with high RTTs, the controller will begin by notifying the low RTTs flows; and this maybe enough to drive the network out of the congestion state.
Therefore, it is clear that even if the root cause of congestion remains, because the high RTTs flows were not notified, it is the low RTT flows that were penalized.
That is why, we added a last mechanism, characterized by Eq.~\ref{eq:equilibrage}.
This mechanism is used to propagate the drop probability from controllers with high frequencies to the ones with low frequencies.

\begin{equation} \label{eq:equilibrage}
  p_{i} = p_{i}*(1-\alpha_{i}) + p_{i-1}*\alpha_{i}
\end{equation}
where \(p_{i}\) and \(p_{i-1}\) are the drop probabilities of categories \(i\) and \(i-1\), respectively.
\(1-\alpha_{i}\) is the decaying factor; it was set to 0.002, 0.004, 0.012, and 0.024 for the categories 2, 3, 4, and 5, respectively.

The frequencies of update of the five categories are \(25 Hz\), \(12.5 Hz\), \(6.25 Hz\), \(3.125 Hz\), and \(1.5625 Hz\), respectively.
This gives a global update frequency of about \(50 Hz\), which is way below the update frequency of most existing controllers.
Knowing this, it becomes clear that even if there are five different controllers running on the router, the computation overhead exerted is not very large.
And with a high performance implementation of the FLC, the computation overhead could be reduced considerably.

Having these five categories is indeed helpful, but the RTTs of the flows can vary greatly, and sometimes it is not clear in which category a flow should fall.
For instance, a flow with an RTT of \(120ms\) is equally distant from the second and third categories.
This brings the question of which drop probability should be used when a flow does not clearly fall in one and only one category.
That is why a second mechanism is introduced which calculates the most appropriate drop probability of the flow using its own RTT and the drop probabilities calculated by the different categories.

\begin{figure}[tb]
  \centering
  \includegraphics[width=3in]{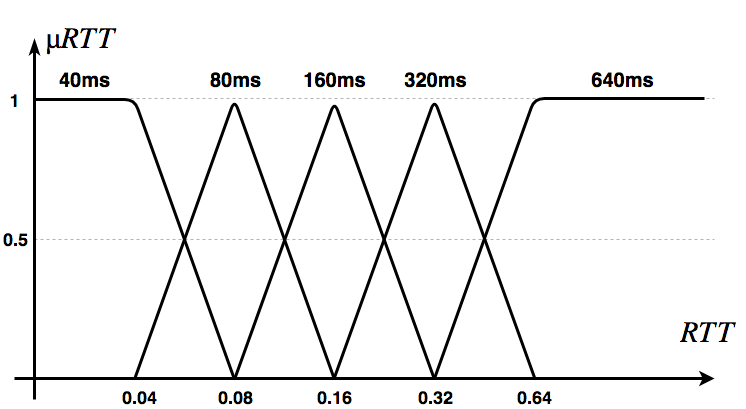}
  \caption{Membership functions of the input variable \(RTT\)}
\label{fig:memFunIn2}
\end{figure}

\begin{figure}[tb]
  \centering
  \includegraphics[width=3in]{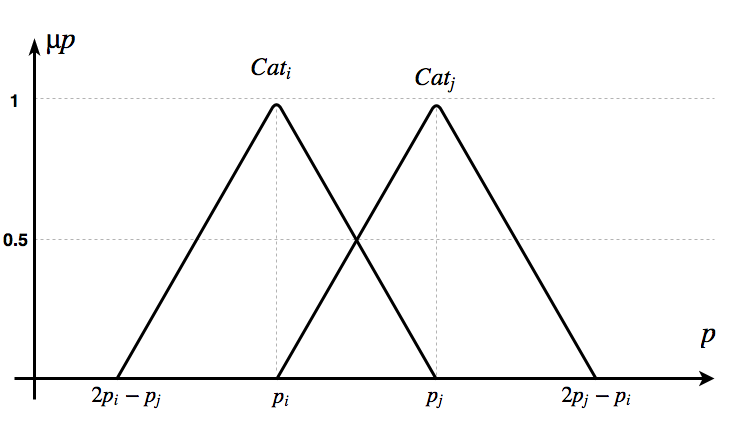}
  \caption{Membership functions of the output variable \(p\)}
\label{fig:memFunOut2}
\end{figure}

The structure of this new mechanism can be considered as a single input single output (SISO) FLC.
As depicted in Fig.~\ref{fig:memFunIn2}, the input has five membership functions, one for each category.
The output has only two membership functions.
Their boundaries are calculated according to the RTT of the received packet.
It is clear that a flow cannot belong to more than two categories at the same time; and, as mentioned previously, the controller is a single input system.
These two reasons combined will ensure that, at most and at any given time, only two membership functions could be active at the same time.
Fig.~\ref{fig:memFunOut2} shows the membership functions of the output.
The crisp value of the output is calculated using the center of gravity.
This mechanism can be characterized by the following equation
\begin{equation} \label{eq:output}
  p = \left[ 1 - \dfrac{(rtt - rtt_i)}{rtt_{i+1} - rtt_{i}} \right] * p_i + \left[1 - \dfrac{(rtt_{i+1} - rtt)}{rtt_{i+1} - rtt_{i}}\right] * p_{i+1} \\
\end{equation}
where \(p\) is the drop probability applied to each packet, \(rtt\) is the RTT of the packet's flow, \(rtt_i\) and \(rtt_{i+1}\) are the RTT of the categories surrounding \(rtt\), and \(p_i\) and \(p_{i+1}\) are the dropping probabilities of the categories surrounding \(rtt\).

Finally, an illustrative example on how the controller operates is given below.
Upon the receiving of the packet, the controller would map the value of the received RTT into at most two linguistic variables.
For instance, given an RTT of \(100ms\), the involved categories would be \(75\% \) of \(Cat2\) and \(25\% \) of \(Cat3\).
Let \(p_{2}\) and \(p_{3}\) be the drop probabilities of \(Cat2\) and \(Cat3\), respectively.
The output would be calculated by
\begin{equation} \label{eq:example}
p = \dfrac{0.75*p_{2} + 0.25*p_{3}}{0.75+0.25}
\end{equation}

\section{Simulation setup}
\label{sec:setup}

\begin{figure}[t]
  \centering
  \includegraphics[width=0.5\textwidth]{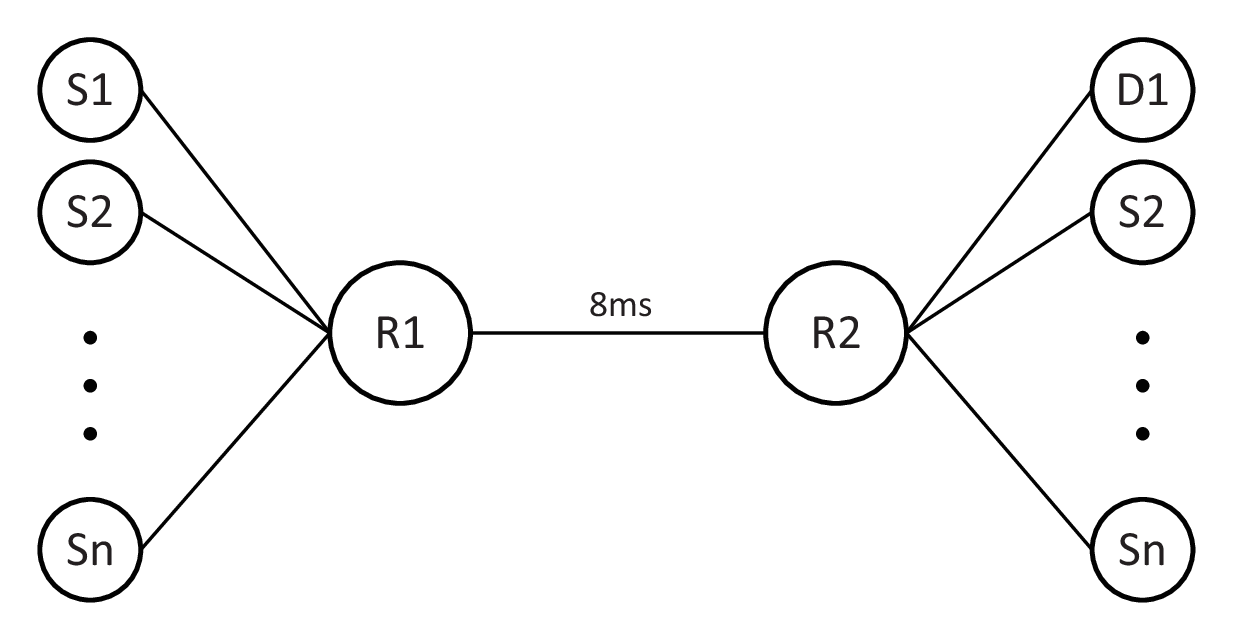}
  \caption{Dumb bell topology}
\label{fig:DumbBellTopo}
\end{figure}

In this section, the simulation setup used to investigate the effectiveness and the robustness of FuzzyRTT is presented.
All the simulations were carried using the OMNeT++ simulator.
In order to demonstrate the effectiveness of FuzzyRTT, a comparison is made against five well-known AQM schemes, which are RED~\cite{Floyd1993}, FLC~\cite{Fengyuan2002}, PID~\cite{Yanfie2003}, CoDel~\cite{Nichols2012}, and CHOKeR~\cite{Lu2014}.
To evaluate the performance of these AQM schemes, a simple dumbbell topology was used (Fig.~\ref{fig:DumbBellTopo}).
This topology is widely used when simulating a long path with a single link bottleneck.
Unless stated otherwise, the following simulation parameters apply.

The delay of the bottleneck link was set to~\(1ms\).
All AQMs are enabled to use ECN except CoDel and CHOKeR.
This is due to the fact that CoDel is designed to drop packets when the minimum queuing time becomes greater than~\(5ms\).
Marking these packets instead of dropping them would lead to the increase of queuing times, which goes against the idea behind the algorithm of CoDel.
While the original CHOKeR drops all packets belonging to the same flow when a hit occurs.
Therefore by not dropping packets when a hit occurs, we shall be changing the behavior of CHOKeR algorithm in a significant manner.
Indeed, during a single RTT, whether a flow receives one or multiple marked packets, it would halve it congestion window only once;
while CHOKeR is designed to free as much space as possible from the queue when a hit occurs.

In all the simulations, the buffer size, or the maximum queue length, was set to the bandwidth delay product.
The TCP flavor used is New-Reno, and the Maximum Transmission Unit (MTU) is set to 576 Bytes (MSS of 536 Bytes).
Finally, the target queue length was set to a queuing time of~\(10ms\).
For instance, with a bottleneck of~\(10Mbps\), the queue target would be \(10\)~packets.

In all the simulations, there are three levels of congestion, light, medium, and high.
When the congestion level is light, the loss ratio is in the vicinity of~\(0.1\% \), and when the congestion levels are medium and high, the loss ratio is in the vicinity of~\(0.5\% \) and~\(1\% \), respectively.
Therefore, the number of FTP flows sharing the link is calculated so as to ensure that the bottleneck link would fall under one of the three levels of congestion.
These three levels of congestion are computed as was described in~\cite{Kuhn2016}.

The fairness between the flows is calculated using the Jain's fairness index.
It is characterized by
\begin{equation} \label{eq:Jain}
J = \dfrac{{(\sum\limits_{i=1}^n x_i)}^{2}}{n\sum\limits_{i=1}^n x_{i}^{2}}
\end{equation}
where \(J\) is the fairness index, \(n\) is the number of flows sharing the link, and \(x_i\) is the throughput of flow \(i\).

\section{Results and discussion}
\label{sec:results}

In this section, the simulation results are discussed.

\subsection{Varying the bandwidth}
\begin{figure*}[!t]
  \centering
  \subfloat[Light congestion]{
    \includegraphics[width=0.33\textwidth]{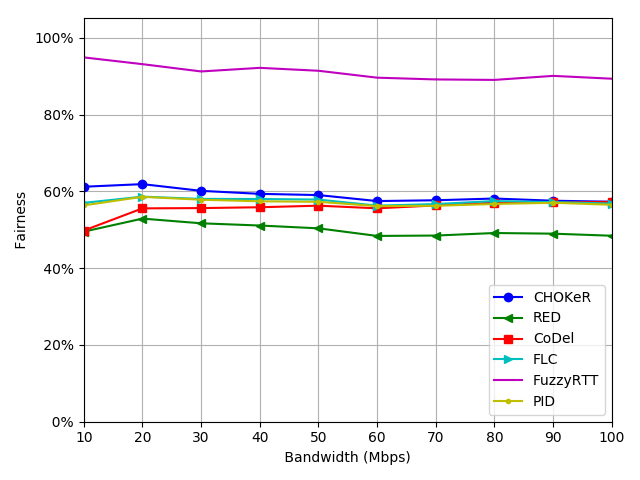}
\label{fig:uni_fair_low}
  }
  \subfloat[Medium congestion]{
    \includegraphics[width=0.33\textwidth]{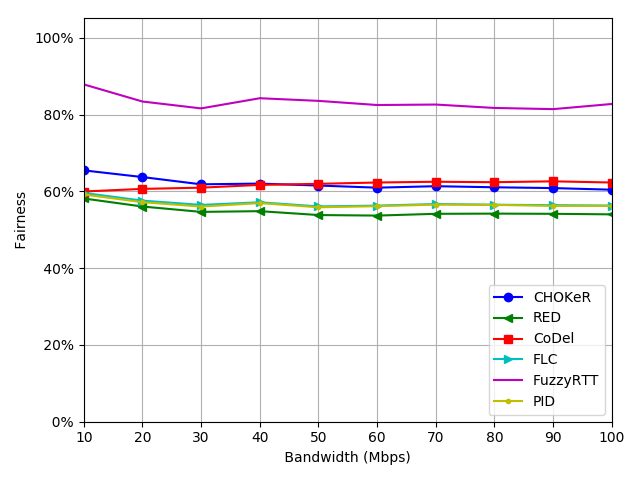}
\label{fig:uni_fair_med}
  }
  \subfloat[Heavy congestion]{
    \includegraphics[width=0.33\textwidth]{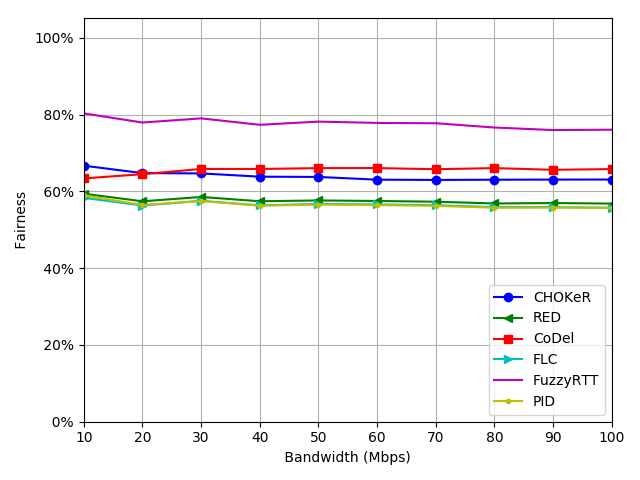}
\label{fig:uni_fair_high}
  }
  \caption{Fairness when the RTTs are uniformly distributed}
\label{fig:uni_fair}
\end{figure*}
In this first set of simulations, the bandwidth of the bottleneck varies from~\(10Mbps\)~to~\(100Mbps\).
The delays of links other than the bottleneck are set so the RTTs would be uniformly distributed between the five categories.

Figure~\ref{fig:uni_fair_low} shows the fairness of the AQM schemes when the congestion is light.
The fairness of FuzzyRTT is above~\(90\% \).
It shows a better fairness of about~\(30\% \) up to~\(40\% \) compared to the other schemes.
When the congestion level is medium (Fig.~\ref{fig:uni_fair_med}), there is a slightly improvement for the other schemes while the fairness of FuzzyRTT drops by~\(7\% \).
But even then, FuzzyRTT is still~\(25\% \) better than the other schemes.
Finally, when the congestion is high (Fig.~\ref{fig:uni_fair_high}), the fairness of FuzzyRTT drops to the vicinity of~\(80\% \).
It still shows a better fairness of about~\(15\% \) compared to the other schemes.

It is clear from Fig.~\ref{fig:uni_fair} that the bandwidth has only a negligible impact on the fairness of all schemes, while the congestion level have a big impact on it.
This is due to the fact that when the bandwidth increases the loss rate stays the same.
Thus, the drop probabilities will be roughly similar.
%
%
Figure~\ref{fig:uni_fair} also shows that when the congestion level increases, the fairness of FuzzyRTT decreases while it increases for the other schemes.
This is due to the fact that all the other schemes use the same drop probability for all passing packets.
Therefore, when the congestion level increases the aggressive flows will have more drops than less aggressive ones.
While, FuzzyRTT protect slow flows, therefore, when the congestion level increases they are forced to drop more and more packets from the slow flows.
It should be noted that even if the fairness of FuzzyRTT drops when the congestion level increases, it is still above~\(80\% \) and there is always a gap of about~\(15\% \) compared to the other schemes.

\begin{figure*}[!t]
  \centering
  \subfloat[Light congestion]{
    \includegraphics[width=0.33\textwidth]{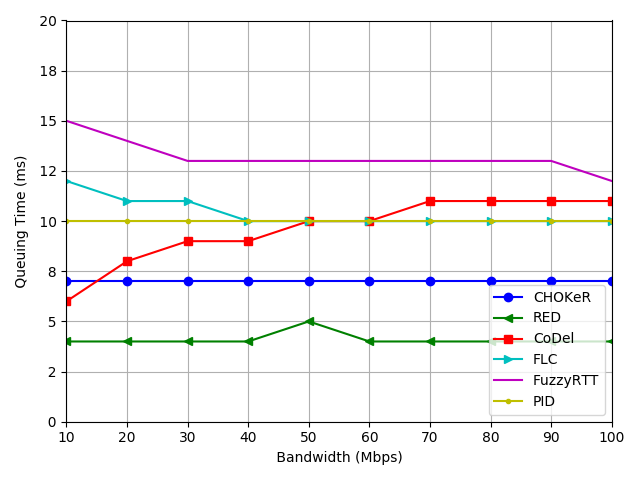}
\label{fig:uni_delay_low}
  }
  \subfloat[Medium congestion]{
    \includegraphics[width=0.33\textwidth]{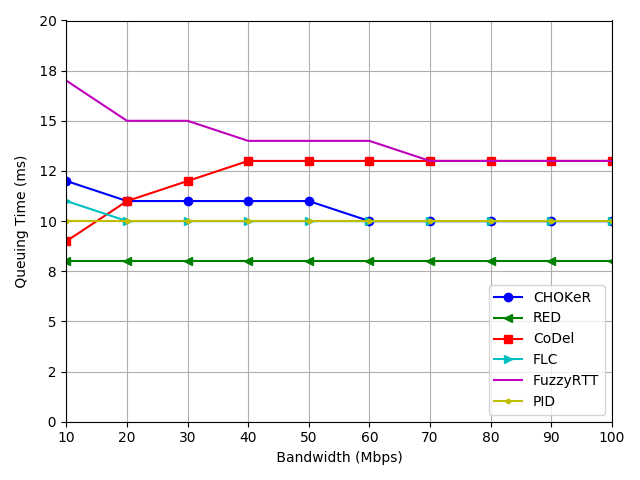}
\label{fig:uni_delay_med}
  }
  \subfloat[Heavy congestion]{
    \includegraphics[width=0.33\textwidth]{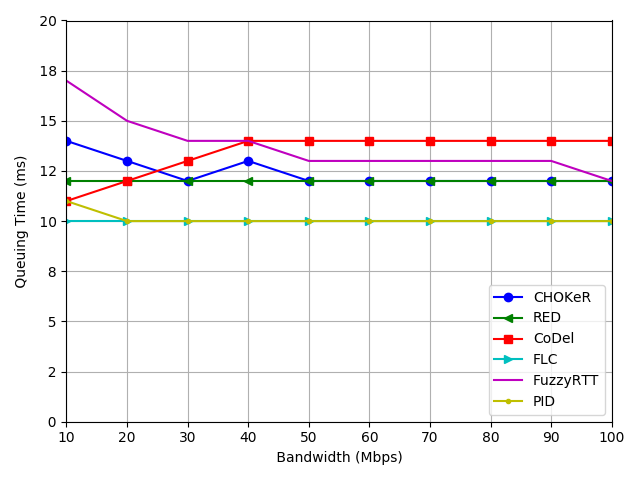}
\label{fig:uni_delay_high}
  }
  \caption{Queuing delay when the RTTs are uniformly distributed}
\label{fig:uni_delay}
\end{figure*}

Figure~\ref{fig:uni_delay} depicts the queuing times of the schemes for all of the three congestion levels.
It can be seen from Fig.~\ref{fig:uni_delay} that the queuing times are roughly stable for all the schemes.
It can also be seen that all the schemes manage to control the queuing delay to be close to the target.
FuzzyRTT, CoDel, and RED show an average error of~\(4ms\), \(2ms\), and~\(-2ms\), respectively, while PID, FLC, and CHOKeR are able to control the queue to the desired target.
The reason why FuzzyRTT is not able to drive the queue to the desired target is mainly due to the low update frequency of the drop probability.
While the other schemes use the same drop probability for all the schemes, FuzzyRTT updates the drop probability of the schemes according to the the flows' RTT\@.
Therefore, it takes more time for FuzzyRTT to adjust its drop probability compared to the other schemes.

\begin{figure}[t]
  \centering
  \includegraphics[width=0.45\textwidth]{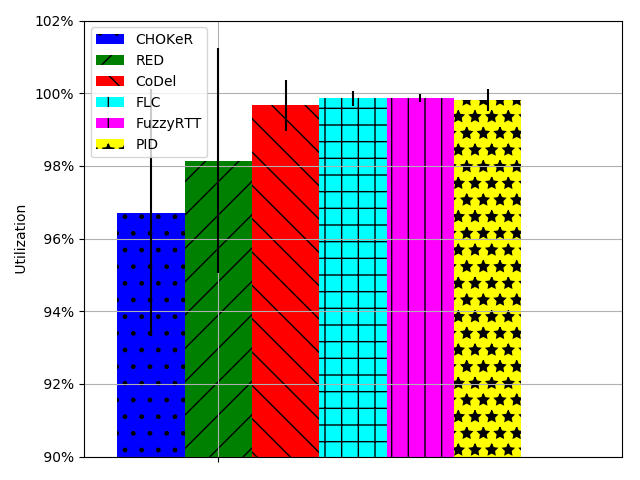}
  \caption{Utilization when the RTTs are uniformly distributed}
\label{fig:BW_utilization}
\end{figure}

Finally, Fig.~\ref{fig:BW_utilization} shows the average and standard deviation of the bottleneck utilization.
All of FuzzyRTT, FLC, PID, and CoDel show near perfect utilization.
The mean utilization for these schemes is above \(99\% \), while the standard deviations are \(0.1\% \), \(0.2\% \), \(0.3\% \), and \(0.7\% \) for FuzzyRTT, FLC, PID, and CoDel, respectively.
On the other hand, the utilization of RED and CHOKeR varies from~\(90\% \), in low bandwidth configurations, and it reaches~\(99\% \) in high bandwidth configurations.

From this set of simulations, it is clear that FuzzyRTT shows both high fairness and utilization but it suffers from a slightly higher queuing delay and jitter compared to all the other schemes.

\subsection{Log normal distribution of RTT}

\begin{figure*}[!t]
  \centering
  \subfloat[Fairness]{
    \includegraphics[width=0.33\textwidth]{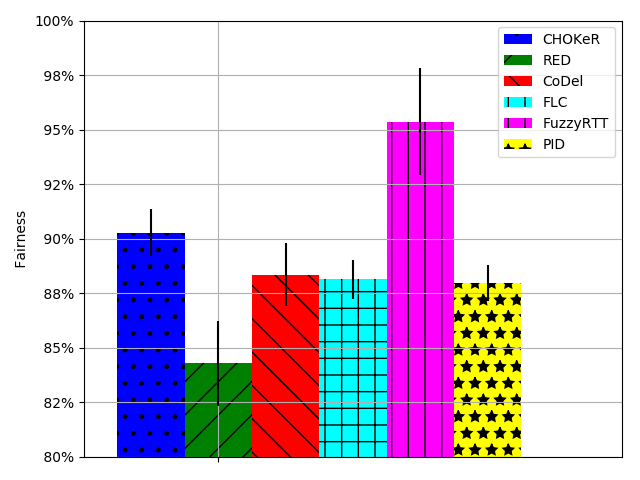}
\label{fig:log_fair}
  }
  \subfloat[Utilization]{
    \includegraphics[width=0.33\textwidth]{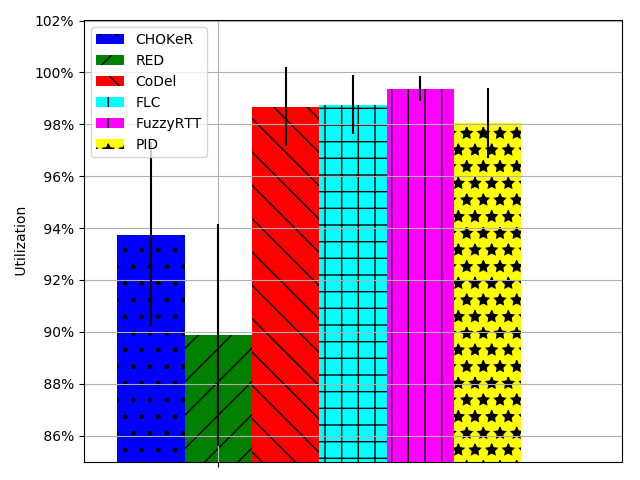}
\label{fig:log_utiliz}
  }
  \subfloat[Delay]{
    \includegraphics[width=0.33\textwidth]{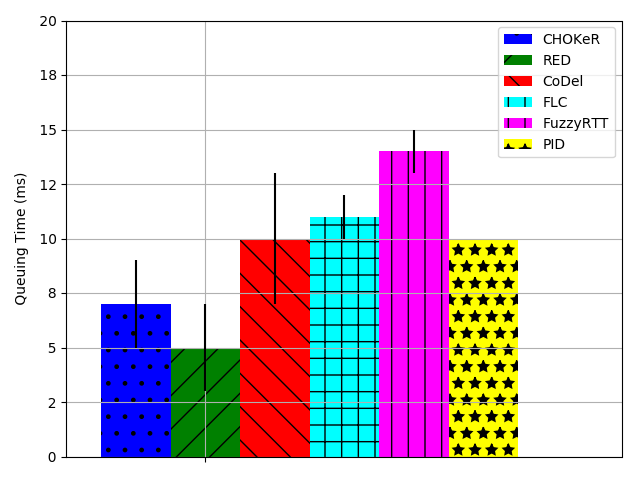}
\label{fig:log_delay}
  }
  \caption{Log normal distribution of RTT}
\label{fig:log_norm}
\end{figure*}

In the Internet, the RTTs of the flows composing the traffic is not uniformly distributed.
According to~\cite{Fontugne2015}, the RTT's distribution is multi-modal which is an aggregation of multiple log-normal distributions.
Therefore, in this set of simulations, the delays of the links are set so the RTTs would follow a log-normal distribution with a mean of~\(\log(149ms)\) and a variation of~\(\log(1.5)\)~\cite{Fontugne2015,Phillipa2006}.
Fig.~\ref{fig:log_norm} aggregates the results of all the simulations for the three levels of congestion where the bandwidth varies from from~\(10Mbps\)~to~\(100Mbps\).
As can be seen on Fig.~\ref{fig:log_norm}, FuzzyRTT shows the best fairness of~\(95\% \) with near optimal utilization.
But as was showed before it cannot drive the queuing time to the target of~\(10ms\).
The second best is CHOKeR with a fairness of~\(90\% \) and an utilization of~\(94\% \).
All the other schemes have a fairness of~\(88\% \) with near optimal utilization, except RED which has a fairness of~\(84\% \) and a utilization of~\(90\% \).

When the distribution of RTTs is log normal most of the flows have RTTs near each other.
Therefore, most of the flows belong to the same category.
Given the fact that all of the proposed schemes show high fairness when the flows have the same RTTs.
Thus, the reason why the gap in fairness between FuzzyRTT and other schemes has dropped is because the share of inter-category fairness have been reduced while the intra-category fairness have been increased.

\subsection{Transient state}

\begin{figure*}[!t]
  \centering
  \subfloat[Delay]{
    \includegraphics[width=0.49\textwidth]{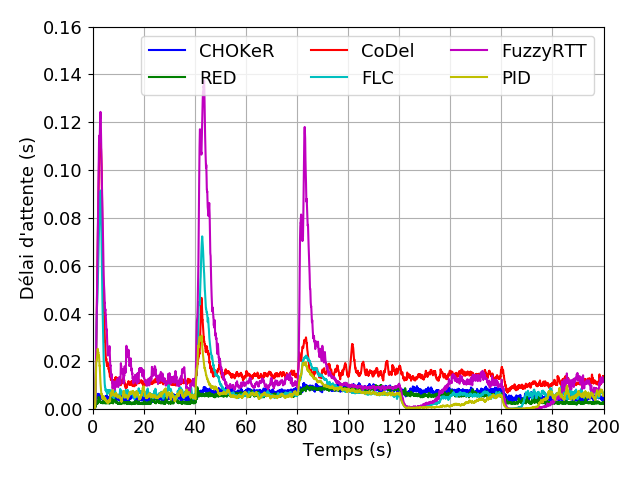}
\label{fig:tran_delay}
  }
  \subfloat[Utilization]{
    \includegraphics[width=0.49\textwidth]{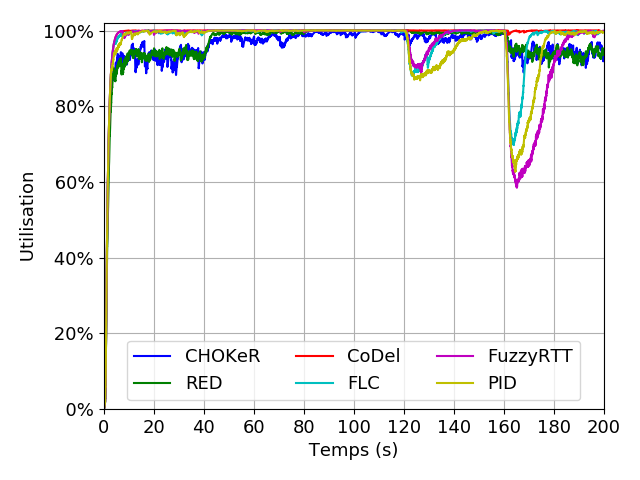}
\label{fig:tran_utiliz}
  }
  \caption{Behavior of AQMs during transient state}
\label{fig:tran}
\end{figure*}

During this simulation exercise, the behavior or AQMs during the transient state is studied.
Doing this will help us compare the time needed by the AQMs to stabilize around the target value.
The bottleneck bandwidth is fixed to~\(50Mbps\).
During the simulation, the bottleneck will undergo three levels of congestion, light, medium, and severe.
Therefore, there are three groups of flows.
When the flows of \(group_1\) are active, the network will be in a state of light congestion, when launching flows of \(group_2\) on top of \(group_1\), the level of congestion will become medium, similarly, the congestion becomes severe when the flows of \(group_3\) become active.
The scenario of this simulation is scripted as follows:
\begin{enumerate}
    \item At \([0s, 2s]\), the flows of \(group_1\) start sending data.
    \item At  \([40s, 42s]\), flows of \(group_2\) start sending data.
    \item At \([80s, 82s]\), the flows of \(group_3\) start sending data
    \item At \([120s, 122s]\), the flows of \(group_3\) stop sending data.
    \item At  \([160s, 162s]\), flows of \(group_2\) stop sending data.
\end{enumerate}

As depicted in \refFig{tran_delay}, FuzzyRTT is the AQM with the longest time to stabilization.
However, the difference between FuzzyRTT, PID and FLC is quite small.
The main reason for which FuzzyRTT shows this lag is due to the low update frequency compared to the other AQMs.
It is clear that RED, CHOKeR and to a lesser extent CoDel, hardly show any transition time when the load exerted on the bottleneck changes.

\refFig{tran_utiliz} shows the utilization of the bottleneck link.
Like shown previously in \refFig{BW_utilization}, all AQMs show a near-optimal utilization except CHOKeR and RED.
When the flows stop sending data, if the drop probability is not updated, the queues will quickly become empty, which will lower the utilization.
Given that the AQMs based on control theory have a relatively small update frequency, they are the AQMs that show the longest time to stabilize.
It is interesting to note that only CoDel can keep high utilization at all times.

It is clear from \refFig{tran} that the reaction of the AQMs is not the same when the congestion changes from light to medium or from medium to light versus when it changes from medium to severe or from severe to medium.
This behavior is more pronounced for FuzzyRTT, PID, and FLC.
The main reason behind this behavior comes from the fact that the number of flows increases/decreases considerably when the congestion changes between light and medium than it does when the congestion changes between medium and severe.
Indeed, the drop rate would go from~\(0.01\% \) to~\(0.05\% \) when the congestion goes from light to medium, while it goes from~\(0.05\% \) to~\(0.1\% \) when the congestion goes from medium to severe.

\section{Deployment considerations}
\label{sec:deployment}

Throughout the paper, as was done in~\cite{Grazia2015}, we assumed that FuzzyRTT was having access to the RTTs of the flows and their number to enforce fairness among them.
In order to run the simulations, as was done with XCP~\cite{Katabi2002}, the TCP header was changed to let the end nodes share their RTTs.
But unlike XCP, FuzzyRTT does not require to actively notify the end nodes about the congestion.
Therefore, any mechanism that can be deployed in the end nodes or in the edge of the network can estimate the RTTs of the flows and append that information to the packets.
By deploying a mechanism such as Explicit RTT Notification (ERN)~\cite{Boudi2018} at the end nodes, we can ensure a high accuracy of RTT estimation.
That mechanism needs only to operate at the IP level and to add the RTT to the IP options header.
The difference between this mechanism and XCP is the fact that by deploying XCP we shall be changing the transport protocol, which will may lead to some issues.
Such as, how can we ensure the stability of a new transport protocol? or how to ensure the friendliness of the new protocol when coexisting with older protocols?
It is clear from past experiences that changing well tested protocols that most users rely upon is indeed difficult.
Because most of the users will not change the protocols unless the newer version grants a high gain compared to the existing ones.
If the RTT information cannot be added to the header of the packets or it is proven to be hard to deploy, FuzzyRTT can still use passive RTT estimation techniques.
But doing this will make FuzzyRTT not a stateless solution anymore.

\section{Conclusion}
\label{sec:conclusion}

With the explosion of the number of connected devices, network overprovisioning may not become a viable option anymore.
Therefore, providing a high degree of fairness among flows while maximizing the throughput of the network nodes will become of utmost importance.
In this paper, we proposed FuzzyRTT, a new Active Queue Management mechanism that aims to enhance fairness among the competing flows.
FuzzyRTT idea is to control the queue by using new information that generally does not exists in routers.
We enabled the end points to share their RTT with routers within the network.
Then, we designed a novel fuzzy logic based controlling mechanism that uses the new available information in order
to enhance the fairness between flows while minimizing queuing delays and improving utilization.
As matter of fact, simulation results showed a better fairness compared to other well-known AQM schemes.
Even if FuzzyRTT was tested under different network conditions, it is not enough to infer that the sensible defaults of its parameters will prove sufficient in real network deployments.
Therefore, in our future work, we aim to study the stability and robustness of FuzzyRTT while also investigating the on-line auto-tuning of its parameters.





\bibliographystyle{IEEEtran}
\bibliography{main}







\end{document}